\documentclass[aps,twocolumn]{revtex4-1}
\usepackage{graphicx}

\begin{document}
\title{Measuring temporal characteristics of the Cherenkov radiation signal \\from extensive air showers of cosmic rays \\with a wide field-of-view telescope addendum to the Yakutsk array}

\author{A.A.~Ivanov}\email{ivanov@ikfia.ysn.ru}
\author{S.V.~Matarkin}
\author{L.V.~Timofeev}

\affiliation{Shafer Institute for Cosmophysical Research and Aeronomy, Yakutsk, 677027 Russia}

\date{\today}

\begin{abstract}
A wide field-of-view Cherenkov telescope has been working in the surroundings of the Yakutsk array experiment since 2012.
Its main function is to measure the waveform of the optical Cherenkov radiation signal
induced by extensive air showers of cosmic rays.
Analysis of the dataset collected by the telescope in the vicinity of $10^{17}$ eV
is intended for the reconstruction of the parameters of the development of the showers
in addition to the main shower characteristics measured by the rest of the array detectors.
In this paper, the observed duration of the Cherenkov radiation signal as a function of the shower core distance is used to estimate the depth of the shower maximum in a different way, based on the results of model simulations. The results are in general agreement with other works.
\end{abstract}

\maketitle

\section{Introduction}
\label{sec:Introduction}
Optical Cherenkov radiation is induced in the atmosphere when a cascade of secondary particles propagate in the air from an initial point where an ultra-high energy (UHE) cosmic ray (CR, in other words, astroparticle) enters the atmosphere. The necessary condition for a charged particle to produce this radiation is to move at a speed greater than $c/n$, where $c$ is the speed of light in vacuum and $n$ is the index of refraction in air~\cite{Tamm}. There are plenty of particles moving at such speeds in an extensive air shower (EAS) of cosmic rays, so Cherenkov radiation can be easily measured on a moonless night with simple light detectors working in coincidence of signals.

Cherenkov radiation is used to infer information about the energy, composition, and direction of arrival of the primary astroparticle that initiated the shower. Since the first observation by Cherenkov \cite{Cherenkov} in the laboratory, and Galbraith and Jelley~\cite{Jelley} in the atmosphere, the systematic measurement of the properties of air Cherenkov radiation were performed in the Pamir experiment~\cite{Chudakov}, and then with a number of EAS arrays. Particularly, the Yakutsk array experiment applies these detectors to estimate the energy and mass composition of the primaries~\cite{Dyak,JETP2007}.

Generally, in previous measurements, analog signal readout systems were used with a narrow bandwidth, restricting the possibility of the reconstruction of the waveform of the Cherenkov radiation signal detected in EAS.
Alternatively, detectors were designed for the measurement of the integral signal \cite{Dyak,Turver,BLANCA}.
However, a digital data acquisition system (DAQ) was recently implemented in the Tunka-133 Cherenkov array, consisting of a set of photomultiplier tubes (PMT)~\cite{Tunka}.
In our papers~\cite{Tmprl,Dcnvlv} a method was described for reconstructing the temporal characteristics of the Cherenkov radiation from the signal of a wide field-of-view (WFOV) Cherenkov telescope (hereinafter `telescope') measured in EAS detected with the Yakutsk array.
These characteristics were used to estimate the parameters of the development of the showers, specifically, to set an upper limit to the
dimensions of the region along the EAS axis
where the Cherenkov radiation intensity is above its half-peak amplitude.

As a development from these efforts, we have analyzed the extended dataset of the telescope measurements including the observational period 2012 to 2015, when coincident detection of EAS events with the telescope was possible.
In the present paper, reconstructed Cherenkov radiation signals are used to estimate in a different way the position of the maximum of the shower particle number in the atmosphere, $X_{\rm max}^{N_e}$, employing the results of Monte Carlo simulations of EAS development.

This article is structured as follows. In Section~\ref{sec:Experiment}, the Yakutsk array experiment and data acquisition and selection for analysis are briefly described, including the telescope. The details of the digital signal processing are given in Section~\ref{sec:SignalProcessing}. In Sections \ref{sec:EAS params} and \ref{sec:Nerling} the connection between the Cherenkov radiation signal and the EAS parameters is studied, and applied to estimate $X_{\rm max}^{N_e}$.

\section{The Yakutsk array. \\Data acquisition and selection for analysis}
\label{sec:Experiment}
\begin{figure}[t]
\includegraphics[width=0.98\columnwidth]{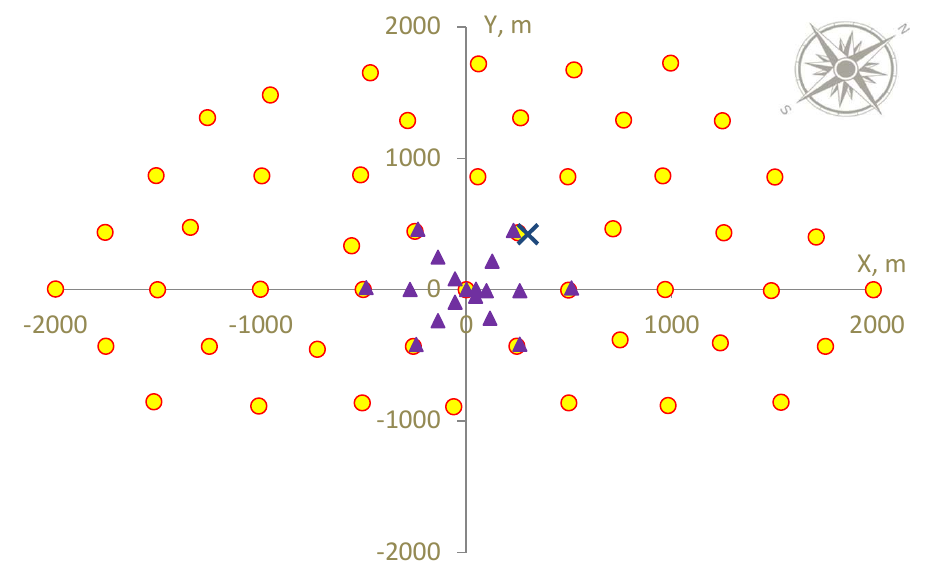}
\caption{The arrangement of the detectors of the Yakutsk array.
Charged particle detectors are shown by open circles, Cherenkov radiation detectors subset by filled triangles.
The position of the telescope is indicated by the cross.}
\label{Fig:Array}\end{figure}

The Yakutsk array is located at a site with geographical coordinates ($61.7^{\circ}N, 129.4^{\circ}E$), 100 m above sea level~\cite{MSU,Zenith}. A schematic view of the layout of the surface stations of the array in the relevant observational period is given in Fig.~\ref{Fig:Array}. Forty-nine stations are distributed within a triangular grid of total area 8.2 km$^2$.
The shower events are selected based on coincidence signals from $n\geq 3$ stations, which in turn have been triggered by the two scintillation counters in each station. Complementary triggers at lower energies are produced by the central cluster consisting of 20 Cherenkov radiation detectors~\cite{KnurCERN,KnurFlor,KnurWeihai}.

The main components of the EAS are detected using scintillators, four muon detectors, 48 air Cherenkov light detectors, and six radio detectors. In this paper, we focus exclusively on the pulse shape of the Cherenkov radiation signal from EAS.
Residual aspects concerning other components of the phenomenon are covered in previous papers of the Yakutsk array group~\cite{JETP2007,Tokyo,EMcomponent,MinWidth,Knur2020}.

All detectors/controllers and data processing units of the array are connected by a fiber-optic network.
An array modernization program aims to achieve a LAN channel capacity of 1 Gbps, synchronization accuracy of detectors, and a time resolution accuracy of $10$ ns. The planned energy range for EAS detection is $(10^{15}, 10^{19})$ eV~\cite{ASTRA,MainResults}.

\subsection{The wide field-of-view telescope detecting the waveform of Cherenkov signals in EAS observed by the Yakutsk array detectors}
\label{sec:Telescope}
\begin{figure}[t]
\includegraphics[width=0.79\columnwidth]{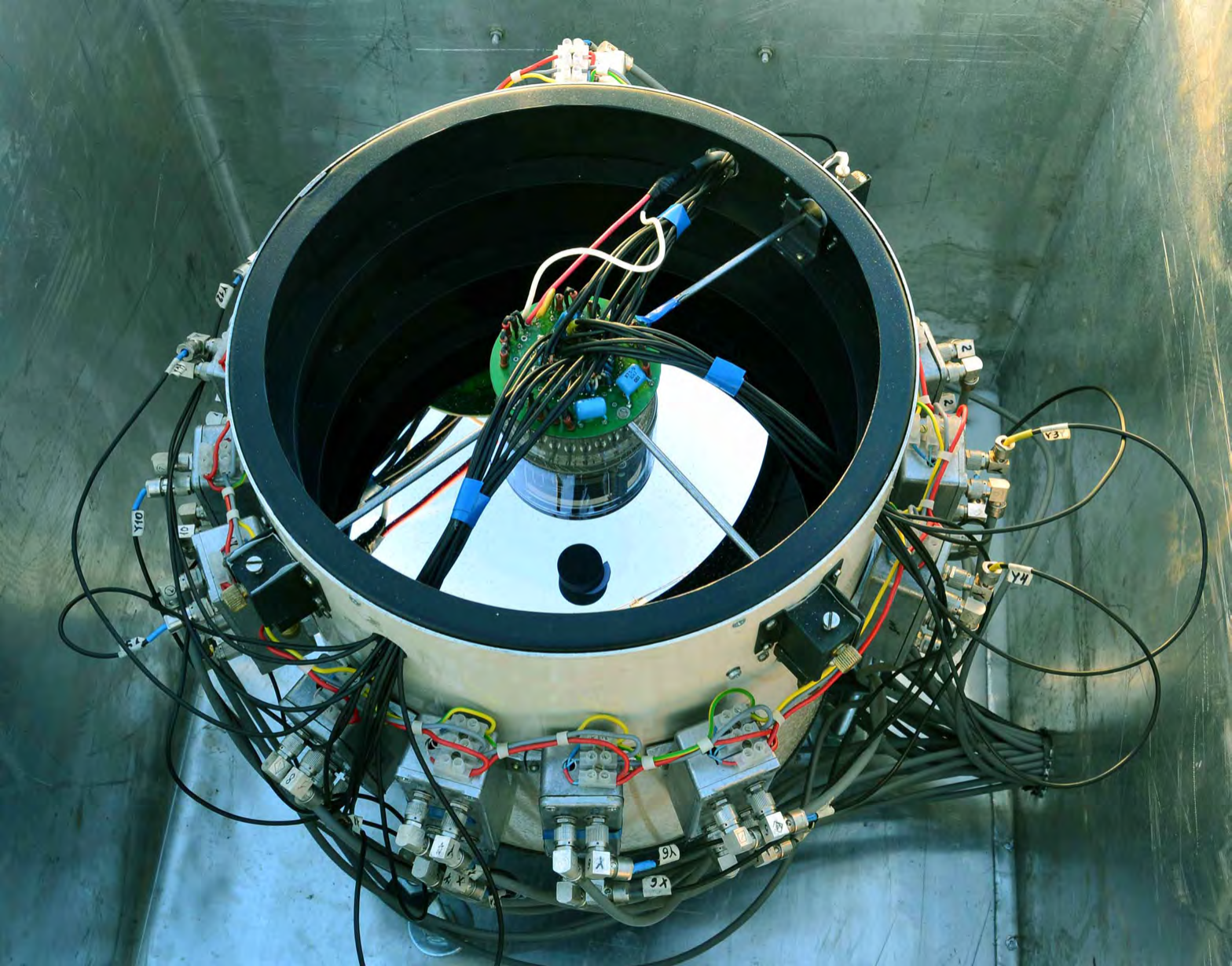}
\caption{Wide field-of-view Cherenkov telescope.
A spherical mirror and a multi-anode PMT with voltage divider and holders are visible.
Pre-amplifiers are mounted on the outside of the tube.}
\label{Fig:Telescope}\end{figure}

The constituent parts of the telescope are a) the spherical mirror (${\o}260$ mm, $f=113$ mm) mounted at the bottom of a metal tube; b) a position-sensitive PMT (Hamamatsu R2486; ${\o}50$ mm) at the focus for which the anode is formed by $16\times16$ crossed wires; c) a voltage-divider circuit and mechanical support attached to the bearing plate; and d) 32 operational amplifiers mounted onto the tube.
The telescope is mounted vertically near an array station (Fig.~\ref{Fig:Telescope}).
A comprehensive description of the telescope can be found in
\cite{ASTRA,Tlscp,Tmprl}.

\begin{figure*}[t]
\includegraphics[width=\textwidth]{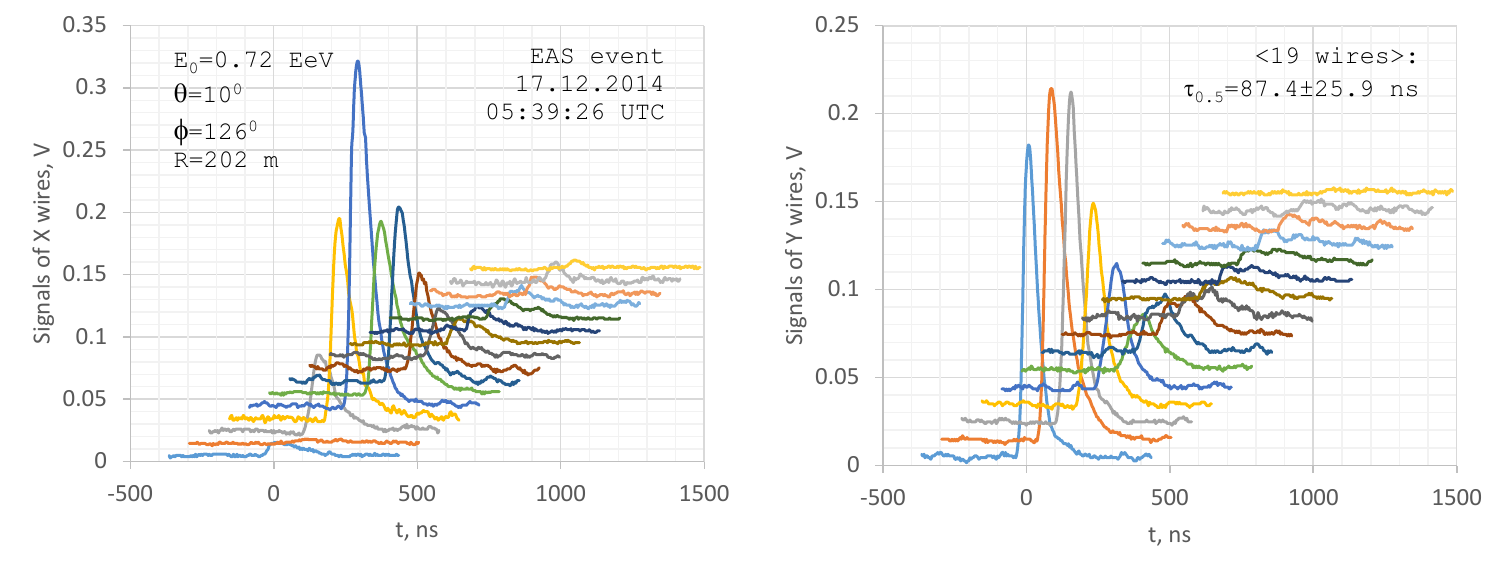}
\caption{The Cherenkov signal from EAS detected with crossed wires of PMT anode of the telescope.
Left panel: 16 X wires; right panel: 16 Y wires.}
\label{Fig:Fout}\end{figure*}

The data acquisition system of the telescope consists of 32 operational amplifiers that have 300-MHz bandwidth AD8055 chips connected by long (12 m) coaxial cables to 8-bit LA-n4USB ADC digitizers with 4-ns time slicing. All of the ADC output signals from the 32 channels are continuously stored in PC memory. A trigger signal from the EAS array terminates the process and signals in a 32 $\mu$s interval preceding a trigger are dumped. In Fig.~\ref{Fig:Fout}, an example is given of the output signals of the DAQ recorded in coincidence with the Yakutsk array detectors in a particular CR shower. EAS parameters are estimated using the data from all the appropriate array detectors.
In this event, nineteen wires of the telescope PMT exhibit significant Cherenkov radiation signals; the other thirteen wires show no signal above the noise level.

In this paper, we use the data accumulated during the period from October 2012 to March 2015 (total number of EAS events is 300173) for which EAS events were detected simultaneously by the surface detectors and the telescope (733 events). Data selection cuts are applied to exclude showers with cores out of the array area and with zenith angles $\theta>60^{\circ}$. The number of EAS events surviving after these cuts is 386.

In the present analysis, we do not use the angular dependence of the telescope signals in an individual EAS event: the angular and arrival time differences of signals are ignored.
Saturated signals (out of 32 wires) in events are ignored, too. The average zenith angle of the showers in the sample is $18^0\pm11^0$, and the energy is $(2\pm0.3)\times10^{17}$ eV.

\section{Analysis of Cherenkov radiation signal}
\label{sec:SignalProcessing}

\subsection{EAS simulation results concerning temporal characteristics of the Cherenkov signal}
\label{sec:CORSIKA}
The physical description of the Cherenkov radiation of relativistic charged particles in a medium originated with the paper of Frank and Tamm \cite{Tamm}. The characteristics of the radiation induced by a cascade of particles in the atmosphere used to be exhaustively modeled by the numerical solution of the cascade equations, or widespread Monte Carlo codes, such as CORSIKA \cite{CORSIKA}. The most known applications of these model simulations are: estimation of the energy of the primary astroparticle using the total flux of the Cherenkov radiation in the EAS \cite{Spectrum_vs_Cher,Spectrum_vs_Gmodel,TotalFlux}; special `Hillas' parametrization of the Cherenkov images of showers in imaging air Cherenkov telescopes, resulting in an unprecedented separation of the very high energy photons, initiating the EAS, from the nuclear background \cite{Hillas}.

The main results of simulations concerning the temporal structure of the Cherenkov signal in EAS are the finding of a near-spherical shower front and that the duration of the signal increases with the shower core distance. To elucidate these features, it is convenient to apply a toy model using a vertical EAS for simplicity. A detector is placed at a distance of $R_i$ far away from the core, so that `the shining point' approximation \cite{ShinePoint} is applicable, namely, the light emitter with normal angular distribution $f_{\rm cher}(\alpha)$, where $\alpha$ is the angle between the direction to the detector and the shower axis, is moving along the shower axis with the speed of light; the light intensity is proportional to the cascade curve, i.e., the total number of electrons, $N_e(h)$.

The photon arrival time to detector is defined by
\begin{equation}
ct=n\sqrt{h^2+R_i^2}-h,
\label{Eq:ShinePoint}
\end{equation}
where $h$ is the emission height; $n$ is the mean refraction index of air; $t=0$ when the shining point arrives at the array plane. For simplicity, we assume here $n=1$ with inaccuracy $\sim 3\times10^{-4}$ \cite{Tlscp}. Integrating $N_e(h)f_{\rm cher}(\alpha)/(h^2+R_i^2)$ one can estimate the total signal of the detector.

The spherical shower front of the photons is evidence that most of the Cherenkov radiation is bounded within a 
small volume around some height $h_{\rm max}^{\rm cher}$. A deviation from sphericity is connected to the width of the cascade curve.

\begin{figure}[b]
\includegraphics[width=\columnwidth]{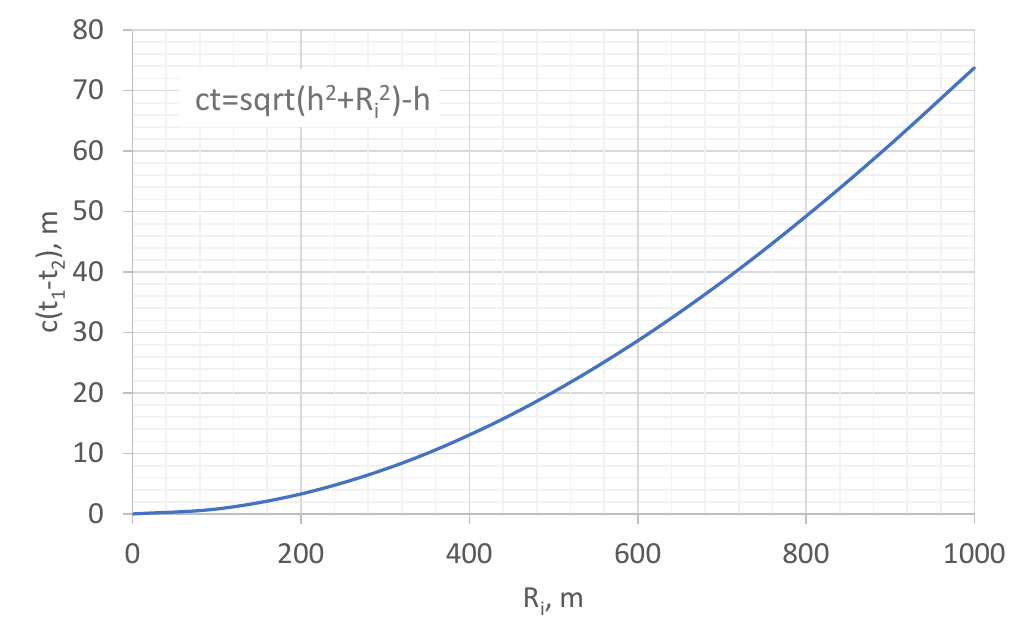}
\caption{Signal duration as a function of the shower core distance, $R_i$, in a toy model with rectangular cascade curve.}
\label{Fig:ToyDur}\end{figure}

The duration of the signal, as a function of the core distance, is produced by a plain geometrical effect which can be demonstrated using Eq.~\ref{Eq:ShinePoint}, specifically,
with the cascade curve of rectangular form, equal to a constant $\neq 0$ between $h_1$ and $h_2$ (Fig.~\ref{Fig:ToyDur}).

\begin{table}[b]
\begin{center}
\caption{Sum of squared residuals of fitted distributions and the waveform functions.
The number of bins are different, so the sums should be compared within columns only.}
\begin{tabular*}{0.48\textwidth}{@{\extracolsep{\fill}}lrr}
\hline\hline
Approximation & Modeled $f_{\rm in}$ & Measured $f_{\rm in}$\\\hline
      Normal  &          11.68   &      -  \\
   LogNormal  &           0.12   &  24.79  \\
       Gamma  &           0.78   &  26.68  \\
\hline\hline
\end{tabular*}
\end{center}
\end{table}

Chitnis and Bhat found \cite{LogNorm} that the waveform of the Cherenkov signal in the detector is represented by a lognormal distribution function fairly accurately at core distances up to 280 m, employing Monte Carlo simulation studies of showers with CORSIKA v.560 and EGS4 codes. Battistoni et al. fitted lognormal and gamma distributions to the delay distributions of secondary photons and electrons for different EAS primaries \cite{LogNorm2}. They conclude that the lognormal distribution fits the data better mainly because of the long tails at large delays ($\sim 200$ ns). We have demonstrated recently \cite{Dcnvlv} that the Cherenkov radiation signal from EAS can be approximated by the gamma distribution, using digital signal processing of the output data from the detector.

\begin{figure*}[t]
\includegraphics[width=0.45\textwidth]{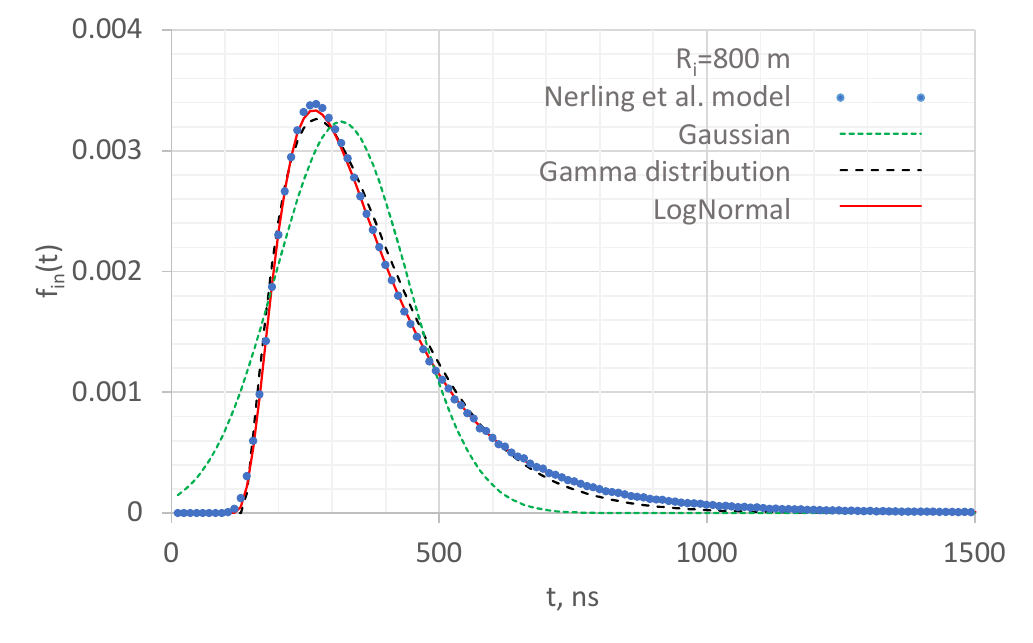}
\includegraphics[width=0.45\textwidth]{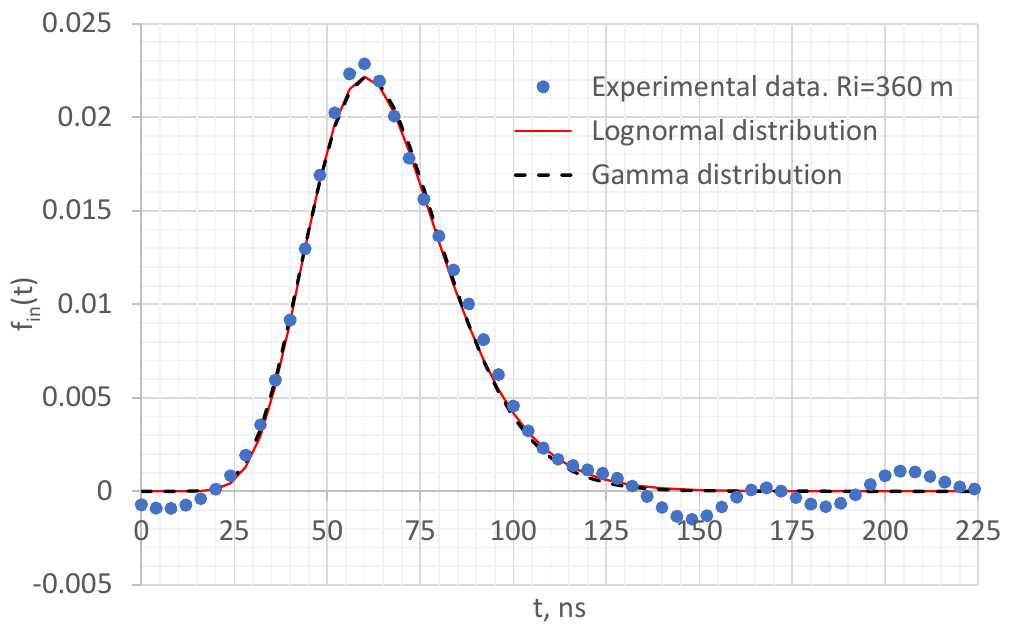}
\caption{Fitting the waveform of the Cherenkov signal in the detector with an appropriate pdf.
Left: $f_{\rm in}(t)$ calculated using Eq.~\ref{Eq:Nerling}.
Right: deconvolution of the telescope signal in EAS detected $23^h48^m00^s$ UTC on 21 October 2012.}
\label{Fig:Fit3}\end{figure*}

In order to prove these results we have chosen a method of calculation after Nerling et al.~\cite{Nerling}, from a multitude of Monte Carlo simulations of EAS development, because of the analytical description of the results concerning Cherenkov radiation in the shower.
The investigation uses an approximation for the energy of the electron and its angular distribution in the high-energy domain based on the universality of both distributions. A similar approach was employed in \cite{Universality,deSouza}. The universality of the calculated electron distributions means their independence from different primary energies, particle types, and zenith angles of EAS to a good approximation for the range of electron energy from 1 MeV to a few GeV, covering the range most important for Cherenkov light emission.

The number of Cherenkov photons arriving at a detector of area $S_d$ at the shower core distance $R_i\gg$ core radius is given by~\cite{ShinePoint,JETP2007}

\begin{eqnarray}
Q_{Sd}\propto\int_0^\infty dh\tau(h)\frac{f_{\rm cher}(\alpha)S_dLcos\theta}{L_d^3}\nonumber\\
\times\int_{E_{th}}^{E_0}dE\frac{dN(h,E,E_0}{dE}\varsigma(1-\frac{E_{th}^2}{E^2}),
\label{Eq:Nerling}\end{eqnarray}

\noindent where $\tau(h)$ is the light transmission coefficient; $L$ is the distance along the shower axis from the shining point to the array plane; $L_d$ is the distance from the shining point to the detector; $f_{\rm cher}(\alpha)$ is the angular distribution of the photons; $dN/dE$ is the electron differential spectrum; $\varsigma(1-\frac{E_{th}^2}{E^2})$ is the number of photons emitted by an electron along 1 g/cm$^2$; and the threshold energy for an electron to emit Cherenkov radiation is 
$E_{th}=\frac{nmc^2}{\sqrt{(n-1)(n+1)}}$.
In this approximation, the photons are assumed to be produced at the shower axis. Solid angle of the detector is defined by the height $L\cos\theta$ and $L_d$.

The parametrization of the electron energy spectrum derived by Nerling et al., $a_0E/(E+a_1)/(E+a_2)$ with constants for fixed shower age $s=3/(1+2X_{\rm max}/x)$ (in the Appendix of~\cite{Nerling}) is used in our calculations. The angular distribution of the Cherenkov photons is approximated by
$$
f_{\rm cher}(\alpha,h,s)=a_s(s)\frac{\exp(-\alpha/\alpha_c(h))}{\alpha_c(h)}+b_s(s)\frac{\exp(-\alpha/\alpha_{cc}(h))}{\alpha_{cc}(h)}
$$
and the parameters are given in the Appendix of~\cite{Nerling}. The angular distribution of photons is a direct consequence of the universal electron angular distribution. The total number of particles as a function of depth is approximated by the gamma distribution (the ``Gaisser--Hillas curve'' used by the PAO collaboration~\cite{GH}) with a depth of the maximum $X_{\rm max}=650$ g/cm$^2$.


The resultant waveform of the Cherenkov signal in a detector placed at the core distance $R_i$ is approximated by normal, gamma and lognormal distributions applying the code ``amoeba,'' which implements the downhill simplex method \cite{Amoeba}, to find the least squares deviation from the input signal, $f_{\rm in}$ (Fig.~\ref{Fig:Fit3}, left panel). For completeness, the experimentally measured (deconvolved) waveform is approximated, too (Fig.~\ref{Fig:Fit3}, right panel). In this case, the input signal to the telescope is reconstructed by applying the Wiener deconvolution algorithm \cite{Dcnvlv}.

It seems that the lognormal and gamma distributions fit the waveform almost equally well, particularly in comparison with the variance of the real signal in the experiment, over the whole range of distances far from the shower core. The sums of the squared residuals are listed in Table 1. Of the three, the lognormal distribution has the minimum deviation, so we have chosen it as the best approximation to the waveform of the Cherenkov signal.

\subsection{Deconvolution of the signal measured with the telescope}
\label{sec:Dcnvlv}
\begin{figure}[t]
\includegraphics[width=0.95\columnwidth]{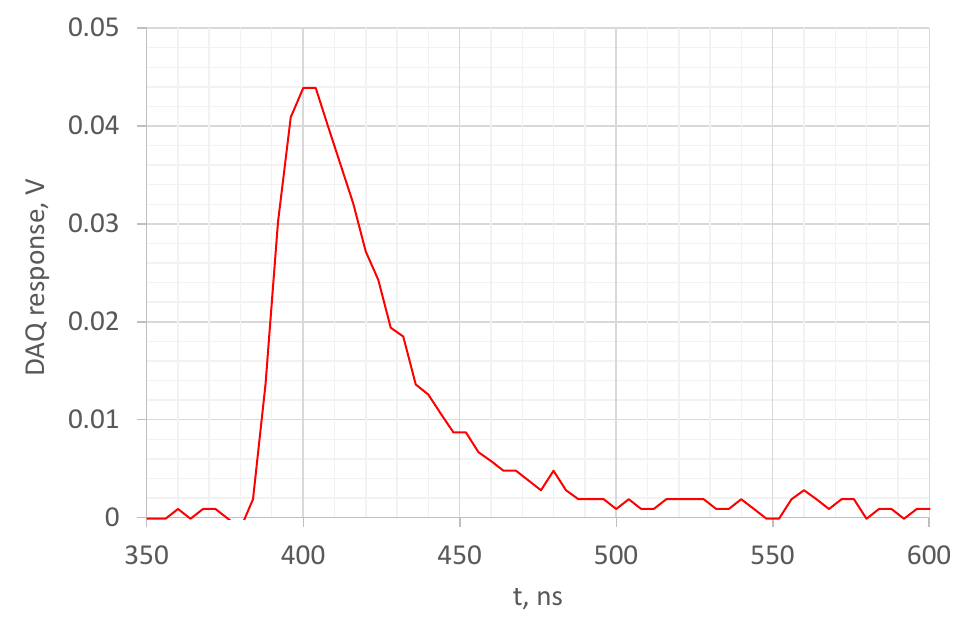}
\caption{Impulse response of the data acquisition system to a short input signal.}
\label{Fig:Response}\end{figure}

The method of deconvolution of the signal observed by the telescope is described in detail in our previous papers~\cite{Dcnvlv,Dcnvlv2}.
In short, an input Cherenkov signal can be reconstructed by the Fourier transform applied to

\begin{equation}
f_{\rm out}(t)=\int_{-\infty}^{\infty} f_{\rm in}(\tau)g(t-\tau)d\tau =(f_{\rm in}*g),
\label{Eq:Cnvltn}
\end{equation}
where $f_{\rm in},f_{\rm out}$ are the input and output signals of the DAQ; and $g(t)$ is a system transfer function \cite{DSP}.
The last is estimated using the dark current impulse of the PMT; an example is given in Fig.~\ref{Fig:Response}.

It is convenient to reconstruct the input Cherenkov radiation signal induced by EAS applying the approximation by the lognormal distribution function $f_{\rm in}^{\rm lognorm}$, discussed above.
In this case the deconvolution procedure can be simplified due to $f_{in}$ restricted within the given kind of function.

Namely, the method now consists of adjustment of the time window to $f_{\rm out}$ and fitting the free parameters of the trial function $f_{\rm in}^{\rm lognorm}$ so that the forward convolution result is congruent to the measured output signal. The convolution theorem ensures that the derived lognormal distribution is the only solution.

To evaluate the free parameters of $f_{\rm in}^{\rm lognorm}$, the nonlinear least squares approach is used. The aim is to minimize the sum of the squared differences between the observed signal and the convolution result in the time window. The optimal values of the parameters are found here by applying a downhill simplex method~\cite{Amoeba}.

To decrease the influence of noise on the analyzed signals, we selected DAQ output signals with amplitudes above the threshold 0.075 V.
It was found sufficient in selection of appropriate signals using the real experimental data of the telescope and the Yakutsk array detectors with signal-to-noise ratio above 45 dB \cite{Dcnvlv}.

\begin{figure}[t]
\includegraphics[width=\columnwidth]{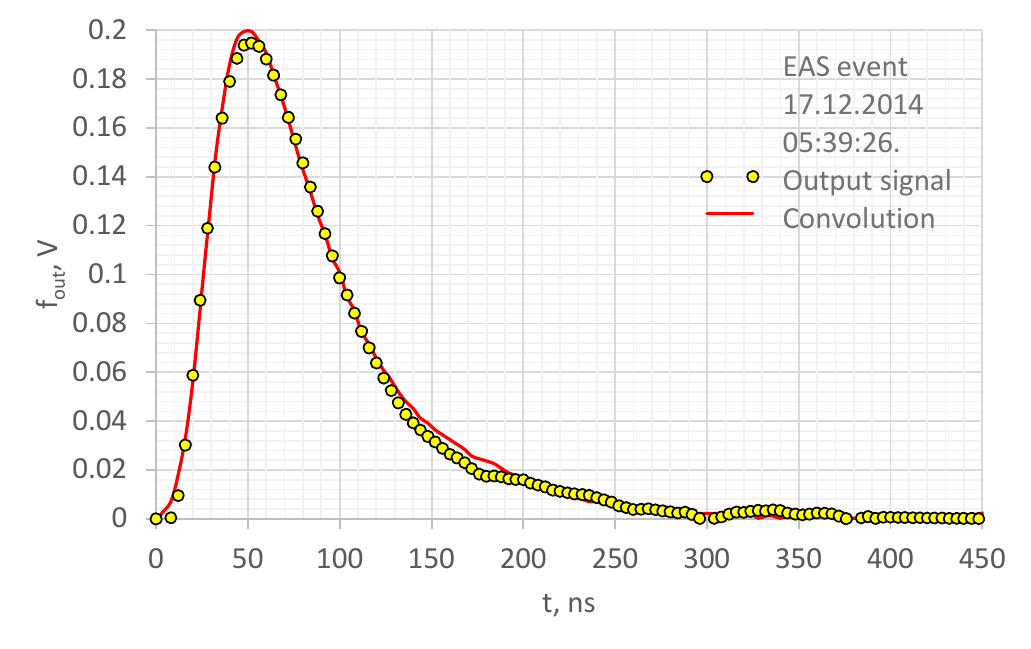}
\caption{The convolution of the impulse response with trial lognormal distribution versus output signal of DAQ.}
\label{Fig:Fit}\end{figure}

For instance, in the event no.~906 shown in Fig.~\ref{Fig:Fout}, only eight channels have amplitudes of the signal above this threshold.
The optimized convolution result $(f_{\rm in}^{\rm lognorm}*g)$ in comparison with the observed output signal $f_{\rm out}$ is illustrated in Fig.~\ref{Fig:Fit}.

The optimization of the lognormal distribution means in our case the fitting of two parameters $av,\sigma$ in order to minimize the sum of squared residuals of the output distributions:

\begin{equation}
f_{\rm in}^{\rm lognorm}(t)=\frac{1}{t\sigma\sqrt{2\pi}}\exp(-\frac{(\ln(t)-av)^2}{2\sigma^2}),
\label{Eq:GammaParams}
\end{equation}
where $av$ is the mean of $\ln(t)$ and $\sigma$ is the rms deviation.

If one has the measured moments of the $t$-distribution: $\bar{t},D_t$, then $\sigma^2=\ln(1+D_t/\bar{t}^2)$; $av=\ln(\bar{t})-0.5\sigma^2$.

\section{Connection of the duration of the Cherenkov radiation signal with the EAS parameters}
\label{sec:EAS params}
As the main parameter of the Cherenkov radiation observed by the telescope, we treat the duration of the signal, i.e., the full width at half maximum, FWHM, of the lognormal distribution recovered from $f_{\rm out}$. It is shown to have a clear dependence on the shower core distance, which can be used to connect it with the development of the EAS in the atmosphere~\cite{Klmkv,Tunka,KnurWeihai,Tmprl}.

Coincident EAS events detected simultaneously with surface detectors of the Yakutsk array and the telescope were selected for analysis.
The shower parameters were estimated based on the data of the surface detectors; signals with amplitudes above the threshold were used from the telescope DAQ channels to infer the average duration of the Cherenkov signals. A bee line from the telescope to the shower axis is used as the core distance of the detector
\begin{equation}
R_i=R_{AP}\sqrt{\sin^2\psi+\cos^2\psi\cos^2\theta},
\label{Eq:Rp}\end{equation}
where $R_{AP}$ is the distance to the core in the array plane, and $\psi$ is the angle between $R_{AP}$ and the projection of the shower axis.

In spite of the additivity of the variance of the signal, we preferred the FWHM because of its ease of use in experiment. Furthermore, it inherits additivity within certain limits. The 32 crossed wires of the anode with private DAQ channels provide at least several independent measurements of a Cherenkov signal above the threshold in an individual EAS event. At another step, showers are selected in the intervals of core distances where the durations of the reconstructed signals are averaged.

\begin{figure}[t]
\includegraphics[width=0.98\columnwidth]{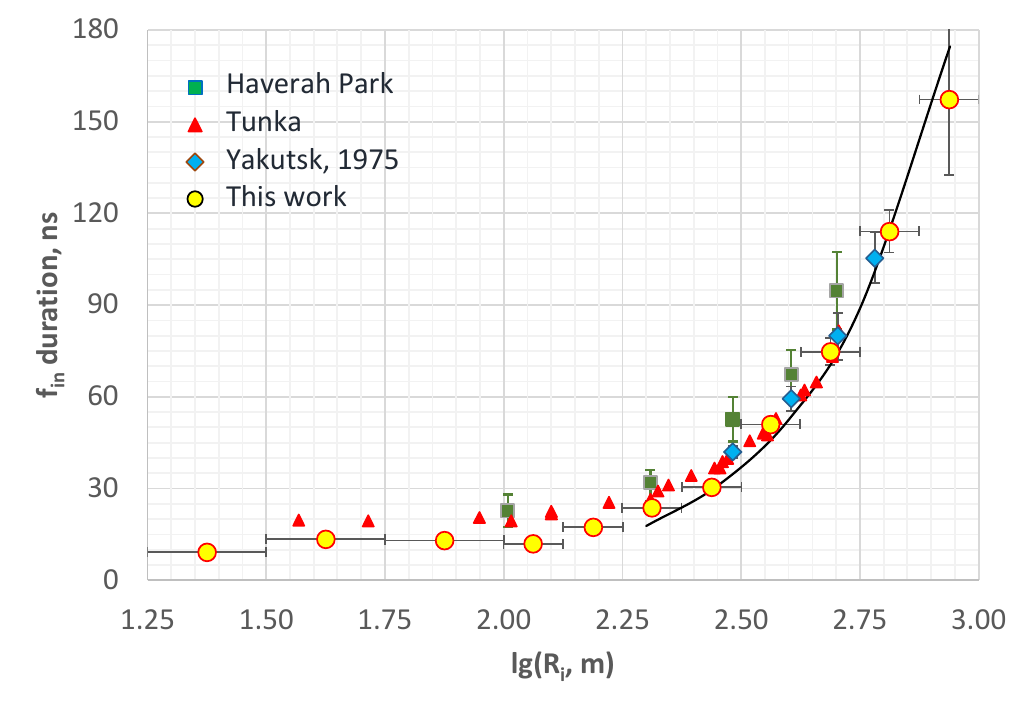}
\caption{Full width of half maximum of the input Cherenkov signal from EAS as a function of the shower core distance. Previous data: Haverah Park \cite{Turver}, Tunka \cite{Tunka}, Yakutsk, 1975 \cite{Klmkv}.
Vertical bars are statistical errors, horizontal bars are intervals of the radial distance.
Solid curve in the interval $R_i\in(200,1000)$ m is the result of the model simulation that will be described in section \ref{sec:Nerling}.}
\label{Fig:FWHM}\end{figure}

The resultant FWHM of the Cherenkov signal measured with the telescope in coincidence with the surface detectors of the Yakutsk array as a function of $R_i$ is shown in Fig.~\ref{Fig:FWHM} in comparison with previous measurements. Our own previous efforts to measure signal durations yielded the results given in~\cite{Tmprl,Dcnvlv}. Since then, the number of measured EAS events has increased, and the reconstruction algorithm has been improved, so the results have become somewhat more enhanced.

The signal duration is almost constant at core distances below $100$ m due to the radius of the shining area in EAS core, and is rising with the radius at $R_i\gg100$ m because of the greater length of the shining area along the axis and the position of the shower maximum in the atmosphere, as was explained in Section \ref{sec:CORSIKA}.

Sampling EAS arrival angles, we have found the function FWHM$(R_i)$ to be independent of the azimuth and zenith angles within instrumental errors. While the independence from the azimuth is not surprising, the zenith angle dependence may be revealed through the distance to $X_{\rm max}^{N_e}$ rising with $\theta$. A possible reason is the insufficiently large aperture of the telescope to reveal a faint zenith angle effect~\cite{Tlscp}.

\begin{figure}[t]
\includegraphics[width=\columnwidth]{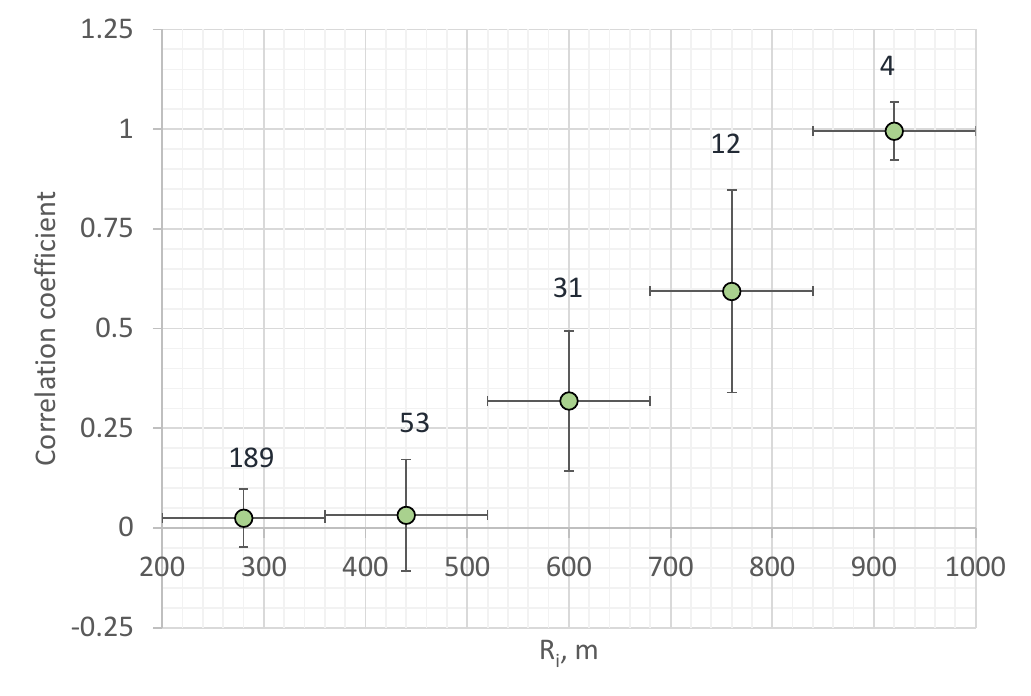}
\caption{The linear correlation coefficient of the signal duration with $lg(E)$ in different shower core distance intervals shown by horizontal bars. Vertical bars indicate statistical errors. EAS event numbers within intervals are placed over data points.}
\label{Fig:DurVsE}\end{figure}

We looked for an energy dependence of the duration of the Cherenkov signal applying Pearson's correlation coefficient
$$
\rho_{x,y}=\frac{cov(x,y)}{\sigma_x\sigma_y},
$$
where the signal duration substitutes x, and $\lg(E)$ substitutes y.
Observational data are sampled in $R_i$ intervals where the linear correlation coefficient is calculated. The results are given in Fig~\ref{Fig:DurVsE}. It seems that a systematic rise of the signal duration with energy is manifested at large distances from the shower core. The mean uncertainty of $\rho_{x,y}$ is estimated as $\delta\rho=\sqrt{(1-\rho_{x,y}^2)/(n-2)}$ for a small sample size $n$.

\section{Application of EAS simulation results to the analysis of Cherenkov radiation signal}
\label{sec:Nerling}
The parameters of the Cherenkov signal, such as the duration of the signal rising with EAS core distance, $\tau(R_i)$, have been used as objects of investigation in a number of experiments. Namely, the SINP MSU group noted that $\tau(R_i)$ at far distances from the shower core is connected with $X_{\rm max}^{N_e}$ calculated in CKP and HMM model simulations~\cite{Klmkv}. Another method of estimating $X_{\rm max}^{N_e}$ was proposed making use of the lateral distribution slope of Cherenkov radiation measured with the Yakutsk array detectors~\cite{MSU}.

The CASA-BLANCA array studied CRs in the energy range 0.3--30 PeV. To find the transformation from the characteristics of the Cherenkov radiation as measured with BLANCA to the depth of shower maximum, the same method as in the Yakutsk array group was used, validated using the CORSIKA simulations with different hadronic interaction models (QGSJET, VENUS, SIBYLL, and HDPM)~\cite{BLANCA}.

In the Tunka experiment, they used two methods of estimating $X_{\rm max}^{N_e}$: the first is based on the shape of the lateral distribution of the intensity of the Cherenkov radiation, just as was the method used in the previous cases; the second uses the sensitivity of the pulse width at the fixed core distance (400 m) to the position of the EAS maximum~\cite{TunkaStatus}.

Using the measured correlation of the duration of the Cherenkov signal with the distance to the shower core, we have estimated an upper limit to
the dimensions of the region along the EAS axis where
the Cherenkov radiation intensity is above the half-peak amplitude~\cite{Dcnvlv}.
The length of the shining volume is found to be
less than 1500 m, and the diameter is less than 200 m in EAS with the primary energy $E_0 = 2.5\times10^{17}$ eV and zenith angle $\theta = 20^0$.

Due to the monotonic relation of the shining point height with the photon arrival time to detector, Eq.~\ref{Eq:ShinePoint}, it is straightforward to estimate $h_{\rm max}^{\rm cher}$ using the time of signal maximum in detector, $t_{\rm max}$, in a model-independent way \cite{Tmprl}:

\begin{equation}
h_{\rm max}^{\rm cher}\sec\theta=\frac{R_i^2-(ct_{\rm max})^2}{2ct_{\rm max}}+R_{AP}\sin\theta\cos\psi,
\label{Eq:Pythagor}
\end{equation}
where $h_{\rm max}^{\rm cher}$ is the height where the Cherenkov radiation is emitted, which forms the maximum of the signal in detector at $R_i$ from the shower core.

Unfortunately, the Yakutsk array in its present configuration is not able to measure the reference arrival time of the shining point to the array plane with sufficient accuracy \cite{Tmprl}. Therefore, an implementation of this promising method should be postponed until the completion of the array modernization program.

\begin{figure}[t]
\includegraphics[width=\columnwidth]{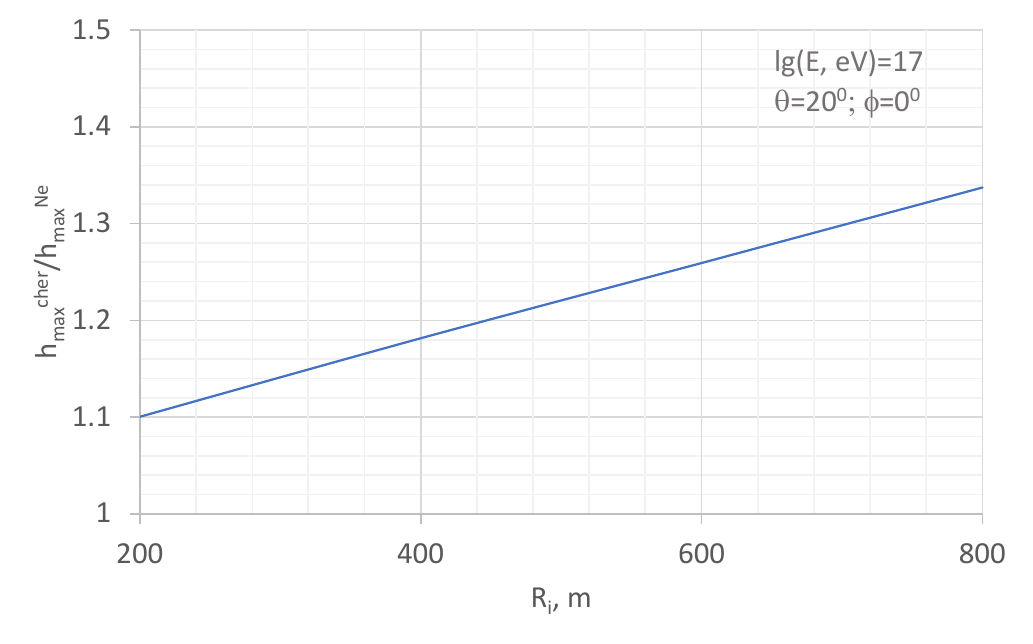}

\caption{The ratio of maximum heights of the Cherenkov radiation intensity and of the number of shower particles calculated in a toy model as a function of the core distance of detector.}
\label{Fig:XmCher}\end{figure}

One of the features of the Cherenkov signal in EAS is that its maximum is different from that of the total number of particles, i.e., the position in the atmosphere of the maximum intensity of the Cherenkov radiation, $h_{\rm max}^{\rm cher}$, is higher than $h_{\rm max}^{N_e}$.
Fig.~\ref{Fig:XmCher} illustrates this property caused by the angular distribution of relativistic electrons emitting Cherenkov photons.
Indeed, evaluation with a toy model indicates that the flat $f_{\rm cher}(\alpha)$ has a weak effect on the position of the maximum, while narrowing the beam leads the visible radiation maximum to drift higher in the atmosphere.

For the purpose of applying the EAS simulation results to the analysis of the Cherenkov radiation signal, namely, to estimate $X_{\rm max}^{N_e}$ based on the temporal characteristics of the Cherenkov signal measured at large shower core distances, it is convenient to employ an analytical description of Cherenkov light emission in EAS, i.e., the results of \cite{Nerling}, as was discussed in Section \ref{sec:CORSIKA}.
Nerling et al.~parametrized the results of the CORSIKA simulations with the QGSJET01 model \cite{QGS01}, which describe showers independently of the primary energy, particle type, and zenith angle, with  a high accuracy of a few percent (within shower-to-shower fluctuations).

Actually, this approach allows one to make use of a toy model with the implemented parametrizations of the CORSIKA simulations.
Adjusting the main unmeasurable parameters of EAS, e.g., $X_{\rm max}^{N_e}$ and the width of the angular distribution of the photons, $\sigma_{\alpha}$, inherent in showers initiated by different nuclei, to measured Cherenkov radiation characteristics, one can find the best fitting values satisfying the conditions of the model.

In general, due to the universality of the electron distributions in EAS, the angular and lateral distributions of the Cherenkov photons emitted by a shower path element depend only on the age of the shower and its height in the atmosphere \cite{UniverCher,UniverAge}.
Consequently, the distributions of photons measured with Cherenkov radiation detectors can be equivalently described by different models having the same $X_{\rm max}^{N_e}$ and $\sigma_{\alpha}$. We do not mention the energy spectrum of electrons, bearing in mind that it determines the total number of electrons emitting Cherenkov radiation, and is parametrized by $X_{\rm max}^{N_e}$.

\begin{figure}[t]
\includegraphics[width=\columnwidth]{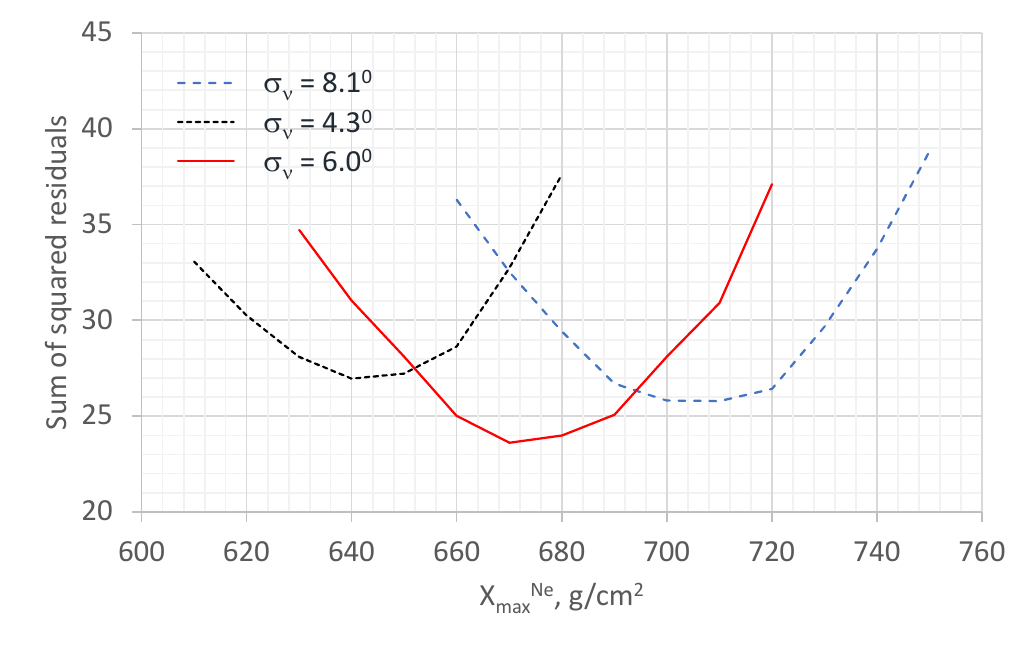}
\caption{Fitting the maximum depth of electrons, $X_{\rm max}^{N_e}$, and angular distribution width of Cherenkov photons, $\sigma_{\alpha}$, in a simulated shower to our observed signal duration as a function of the core distance within $(200,1000)$ m, illustrated in Fig.~\ref{Fig:FWHM}.}
\label{Fig:DurVsXmax}\end{figure}

The difference between this algorithm and that of SINP MSU and Tunka's second approach is in the fixed core distance of $\tau(R_i)$ in their case, and on the contrary, a variety of distances (within the interval $200-1000$ m) in our case, to determine $X_{\rm max}^{N_e}$.
In the latter, the amount of empirical information is definitely greater.

Fig.~\ref{Fig:DurVsXmax} presents a fit to our measurements of calculations with different maximum depths incorporated into the approximations of Nerling et al. by CORSIKA simulations in $R_i\in(200,1000)$ m. Namely, the sum of squared differences between the observed and simulated durations of Cherenkov signals in the $R_i$ interval is minimized. The width of the angular distribution of Cherenkov photons in EAS is a function of the age of the shower and the height of the shining point; we have approximated it by the value at $s=1,h=h_{\rm max}^{\rm cher}$ in order to demonstrate a fit. Variation of the width,
$\sigma_{\alpha}(s,h)$, within the interval $(4.3^0,8.1^0)$ is carried out by a scaling factor applied to the angle between the direction to the detector and the shower axis.
In other words, we have adapted the toy model parameters in order to get the best description of the observed FWHM$(R_i)$ in the interval $R_i\in(200,1000)$ m.


It turns out that $X_{\rm max}^{N_e}=670\pm20 \pm5$ g/cm$^2$ provides the best fit to the experimental values of the durations of the Cherenkov signals in EAS with energy 0.2 EeV and zenith angle $18^0$. Specifically, the optimized model result is in agreement with an overview formed by 386 EAS events allocated to $R_i$ intervals. 
Here, we considered the two sources of $X_{\rm max}^{N_e}$ uncertainties originated in reconstruction errors of the shower core position and arrival direction, and in uncertainties of the Cherenkov signal measurement
(details of the estimation of instrumental uncertainties are given in Appendix \ref{app-a}).

A comparison of the results of measurements of Cherenkov radiation with the simulations in a toy model employing the fitted parameters is given in Fig.~\ref{Fig:FWHM}.

We have chosen here the depth of the shower maximum, $X_{\rm max}^{N_e}$, as a conventional parameter useful for comparison with other experiments.
At the site of the Yakutsk array, for the Cherenkov radiation measurements in winter nights, when the atmosphere temperature profile is close to isothermal, a plain exponential equation $X=\rho_0 h_{\rm atm}\exp(-h/h_{\rm atm})$, where $h_{\rm atm}=7100$ m; $\rho_0$ is air density at $h=0$, can be applied for estimations with inaccuracy $\sim 1\%$ (Appendix \ref{app-b}).

\begin{figure}[t]
\includegraphics[width=\columnwidth]{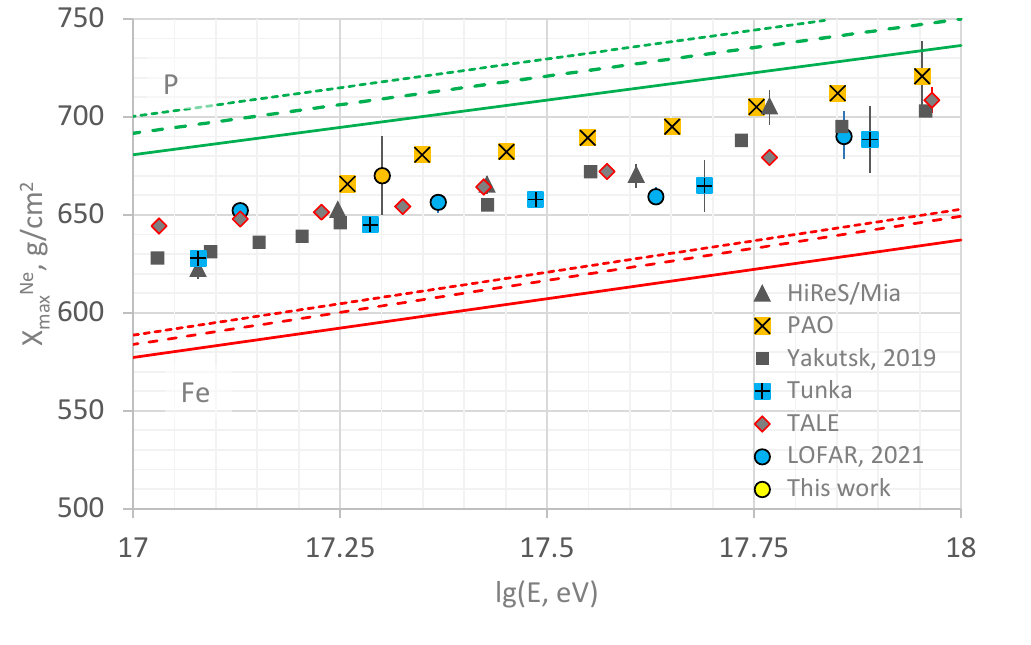}
\caption{World data on $X_{\rm max}^{N_e}$ estimations in EAS collected in \cite{LOFAR} with the present result added.
The lines indicate the simulation results with QGSJetII-04 (solid), EPOS-LHC (dashed) and Sibyll-2.3d (dotted) models for iron and proton primaries.}
\label{Fig:XmaxData}\end{figure}

A resultant average depth of shower maximum in the number of EAS particles is compared with previous measurements in Fig.~\ref{Fig:XmaxData} borrowed from \cite{LOFAR}. It is in reasonable accord with a set of experiments: HiRes \cite{HiRes}, PAO \cite{PAO}, TALE \cite{TALE}, Tunka \cite{TunkaStatus}, Yakutsk 2019 \cite{KnurXmax}, LOFAR \cite{LOFAR} within the interval $(1.7-2.3)\times10^{17}$ eV where the depth dispersion is confined to $\sim(640,680)$ g/cm$^2$.

The estimated mean value of $X_{\rm max}^{N_e}$ at $\overline{E}=0.2$ EeV can be used to infer the proton component fraction in the primary beam, within the two-component (H and Fe nuclei) mass composition assumption. Taking into account the shower maximum depths derived in the QGSJetII-04 \cite{QGSJet}, EPOS-LHC \cite{EPOS}, and Sibyll-2.3d \cite{Sibyll_a,Sibyll_b} models, one concludes that the proton fraction is $79\pm21\%$, $62\pm19\%$, and $56\pm18\%$, and the mean mass $\overline{lnA}$ is 0.85, 1.53, 1.76 at $E=0.2$ EeV, in the corresponding model. These values are close to the results of PAO \cite{PAO}.

The divergence between the estimated values of $X_{\rm max}^{N_e}$ in experiments can be considered to be caused by the model uncertainties and instrumental errors due to the variety of detectors used, from fluorescent and Cherenkov light detectors to radio wave receivers.
A straightforward way to reduce the uncertainties would be the application of model-independent methods of measurement. Regarding the planned Cherenkov radiation measurement in the EAS investigation, the triangulation method employing the shower front curvature, e.g., Eq.~\ref{Eq:Pythagor}, seems to be the best choice.

\section{Conclusions}
The addition of a wide field-of-view telescope to the multitude of the Yakutsk array detectors has expanded its possibilities for EAS investigation for the measurement of the temporal characteristics of the Cherenkov radiation emitted by shower particles.
In the present paper, the results of an enhanced analysis of the temporal features of this radiation detected in coincidence of signals by the telescope and surface detectors have been given.

The input signal of the telescope's DAQ is reconstructed applying a lognormal approximation of the Cherenkov radiation signal from EAS: both measured and simulated by a model. The experimental data are deconvolved from the telescope output signal using an independent method.
The resultant Cherenkov signal reconstruction algorithm is simple and fast, allowing on-the-fly analysis of measured signals.

The main measurable temporal characteristic of Cherenkov radiation induced by EAS is the signal duration. We have enhanced previous measurements of the signal duration and confirmed explicitly that it rises with the shower core distance at $R_i>200$ m. This rise is related to the development of the shower in the atmosphere, and further, we have demonstrated that the behaviour of the signal duration in the interval $R_i\in(200,1000)$ m can be used to estimate $X_{\rm max}^{N_e}$.

An essential requirement for this is the application of EAS modeling under certain assumptions concerning interactions of the particles.
We implemented Monte Carlo simulation results after Nerling et al.~\cite{Nerling} in our toy model calculations. The resultant estimation of the shower maximum depth $X_{\rm max}^{N_e}=670\pm20\pm5$ g/cm$^2$ at $E=(2\pm0.3)\times10^{17}$ eV, $\theta=18^0\pm11^0$ is in reasonable agreement with previous results obtained using different experimental techniques. The connected estimation of the proton fraction and of the mean mass of the primary astroparticles under the two-component hypothesis is close to the results of the PAO collaboration.

\acknowledgements
We would like to thank the Yakutsk array group for data acquisition and analysis. This work was supported by the Ministry of Science and Higher Education of the Russian Federation (program ``Unique Scientific Installations,'' no. 73611).

\appendix
\section{Estimation of experimental uncertainty in $X_{\rm max}^{N_e}$ reconstruction}
\label{app-a}
In our experiment, EAS events detected with the telescope and surface detectors of the array in coincidence of signals provide the mean Cherenkov signal duration in a set of the shower core distance intervals. The toy model employing results of Monte Carlo simulations of EAS development \cite{Nerling} can be adjusted to experimental data by selection of model parameters, e.g. $X_{\rm max}^{N_e}$.

\begin{figure}[t]
\includegraphics[width=\columnwidth]{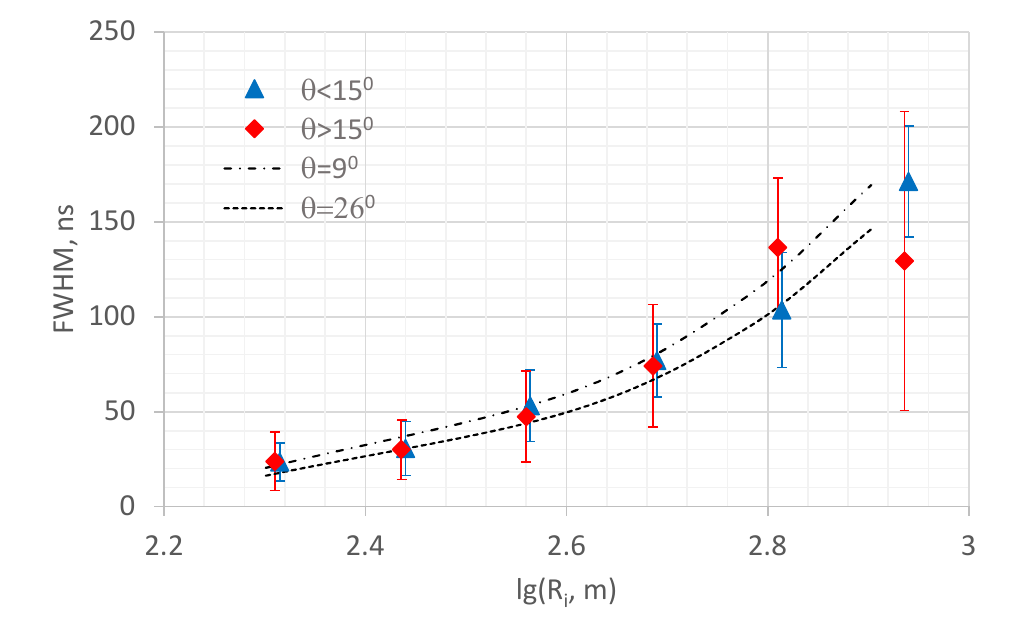}
\caption{Measured Cherenkov signal duration, FWHM, as a function of $R_i$ in two zenith angle intervals separated by $\theta=15^0$. Vertical bars are statistical errors. Two curves are model results calculated for the mean angles in the intervals.}
\label{Fig:HalfEvents}\end{figure}

Systematic uncertainties and statistical errors of observed values can be visualized by the data set divided into two $\theta$ bins of congruous shower samples comparing the signal durations obtained (Fig. \ref{Fig:HalfEvents}). A conclusion to be drawn is that statistical uncertainties in this particular case are greater then expected divergence due to different zenith angles in data samples.

Experimental uncertainties in shower core and arrival direction reconstruction lead to uncertainties of the detector distance to shower axis, $R_i$, and Cherenkov signal duration, $\tau(R_i)$, which in turn result in the uncertainty of fitted $X_{\rm max}^{N_e}$. To estimate an upper limit to the shower maximum depth uncertainty, we have used the toy model with a set of $X_{\rm max}^{N_e}$ within the interval, $\delta X_m$, resulting in the dispersion of $\tau(R_i)$ below that value which is distinctive to the sum of shower core and arrival direction uncertainties.
For the latter we assumed the shower core, $\pm50$ m, and arrival direction, $\pm5^0$, uncertainties \cite{Mono,Pravdin}. The resulting upper limit to the experimental uncertainty is $\delta X_m=5$ g/cm$^2$.

A confidence interval of fitted $X_{\rm max}^{N_e}$ due to experimental and statistical errors (including instrumental uncertainty due to DAQ electronics) is found to be $\pm 20$ g/cm$^2$ at the $95\%$ level assuming a small sample of equiprobable depths.

\section{International Standard Atmosphere}
\label{app-b}
The International Standard Atmosphere (https://www.iso.org/standard/7472.html) has been established to provide a common reference for parameters of the Earth's atmosphere. It consists of tabulated values of temperature, $T$, at 7 altitudes, $h$, that should be linearly interpolated between. Air in the model is assumed to be dry and clean and of constant composition. Neither does it account for humidity effects.

Assuming hydrostatic balance $\frac{dP}{dh}=-g\rho$, where $g$ is gravitational acceleration,
and ideal gas with equation $P\propto\rho T$, the barometric pressure can be calculated $$P=P_0\exp(-\frac{I(h)}{h_{\rm atm}})$$
and the density of air
$$\rho=\rho_0\frac{T_0}{T(h)}\exp(-\frac{I(h)}{h_{\rm atm}}),$$
where $I(h)=\int_0^h\frac{T_0 dz}{T(z)}$; values subscripted $0$ are at $h=0$.

The depth of the atmosphere above $h$ is
$$x=\rho_0 h_{\rm atm}\int_h^\infty\frac{T_0}{T(h)}\exp(-\frac{I(h)}{h_{\rm atm}})\frac{dh}{h_{\rm atm}}.$$ It is defined by the temperature profile of the atmosphere at the site, mainly by a temperature lapse rate in the troposphere. In our case, Polar temperature profile is appropriate with the mean winter night temperature $T_0=243^0\pm10^0$ K and pressure $P_0=1006\pm6$ hPa \cite{Mono,Seasonal}.

There is a simplified formula in a model of isothermal atmosphere
$$x=\rho_0 h_{\rm atm}\exp(-h/{h_{\rm atm}})$$ assuming uniform temperature $T(h)=T_0$. A difference in $x$ values calculated in the two models at $h>3$ km is less than $1\%$ for the same temperature $T_0$. On the contrary, considerable divergency of the air density and the depth of the atmosphere arises in both models due to seasonal and diurnal variations of the temperature and pressure. To account for these changes, $T_0$ and $P_0$ are measured and recorded in each shower event when the Yakutsk array detector signals are triggered \cite{Mono}.


\end{document}